\documentstyle[11pt,epsf]{article}
\begin{document}
\catcode`@=11
\long\def\@caption#1[#2]#3{\par\addcontentsline{\csname
  ext@#1\endcsname}{#1}{\protect\numberline{\csname
  the#1\endcsname}{\ignorespaces #2}}\begingroup
    \small
    \@parboxrestore
    \@makecaption{\csname fnum@#1\endcsname}{\ignorespaces #3}\par
  \endgroup}
\catcode`@=12
\newcommand{\newc}{\newcommand}
\newc{\gsim}{\lower.7ex\hbox{$\;\stackrel{\textstyle>}{\sim}\;$}}
\newc{\lsim}{\lower.7ex\hbox{$\;\stackrel{\textstyle<}{\sim}\;$}}
\newc{\mtpole}{M_t}
\newc{\mbpole}{M_b}
\newc{\mqpole}{m_q^{\rm pole}}
\newc{\mlpole}{m_l^{\rm pole}}
\newc{\MS}{{\rm\overline{MS}}}  \newc{\msbar}{\MS}
\newc{\DR}{{\rm\overline{DR}}}
\newc{\tanb}{\tan\beta}
\newc{\stopq}{{\widetilde t}}
\newc{\stopone}{\widetilde t_1}
\newc{\stoptwo}{\widetilde t_2}

\newc{\mstopq}{m_{\tilde t}}
\newc{\mstopl}{m_{\tilde t_L}}
\newc{\mstopr}{m_{\tilde t_R}}
\newc{\mstopone}{m_{\tilde t_1}}
\newc{\mstoptwo}{m_{\tilde t_2}}
\newc{\zbb}{Z\to b\bar}
\newc{\Gb}{\Gamma (Z\to b\bar b)}
\newc{\Gh}{\Gamma (Z\to {\rm hadrons})}
\newc{\xci}{m_{\chi^{\pm}_i}}
\newc{\xcj}{m_{\chi^{\pm}_j}}
\newc{\xck}{m_{\chi^{\pm}_k}}
\newc{\xni}{m_{\chi^{0}_i}}
\newc{\xnj}{m_{\chi^{0}_j}}
\newc{\xnk}{m_{\chi^{0}_k}}

\newc{\sti}{m_{\tilde {t}_i}}
\newc{\stj}{m_{\tilde {t}_j}}
\newc{\stk}{m_{\tilde {t}_k}}
\newc{\sbi}{m_{\tilde {b}_i}}
\newc{\sbj}{m_{\tilde {b}_j}}
\newc{\sbk}{m_{\tilde {b}_k}}

\newc{\cw}{\cos\theta_W}
\newc{\sw}{\sin\theta_W}
\newc{\mhp}{m_{H^\pm}}
\newc{\mA}{m_{A^0}}

\def\NPB#1#2#3{Nucl. Phys. {\bf B#1} (19#2) #3}
\def\PLB#1#2#3{Phys. Lett. {\bf B#1} (19#2) #3}
\def\PLBold#1#2#3{Phys. Lett. {\bf#1B} (19#2) #3}
\def\PRD#1#2#3{Phys. Rev. {\bf D#1} (19#2) #3}
\def\PRL#1#2#3{Phys. Rev. Lett. {\bf#1} (19#2) #3}
\def\PRT#1#2#3{Phys. Rep. {\bf#1} (19#2) #3}
\def\ARAA#1#2#3{Ann. Rev. Astron. Astrophys. {\bf#1} (19#2) #3}
\def\ARNP#1#2#3{Ann. Rev. Nucl. Part. Sci. {\bf#1} (19#2) #3}
\def\MODA#1#2#3{Mod. Phys. Lett. {\bf A#1} (19#2) #3}
\def\ZPC#1#2#3{Zeit. f\"ur Physik {\bf C#1} (19#2) #3}
\def\APJ#1#2#3{Ap. J. {\bf#1} (19#2) #3}
\def\beq{\begin{eqnarray*}}
\def\eeq{\end{eqnarray*}}
\def\bea{\begin{eqnarray*}}
\def\eea{\end{eqnarray*}}
\newc{\rb}{R_b}			\newc{\rbmax}{\rb^{\rm max}}
\newc{\mtop}{m_t}		\newc{\mtopmax}{{\mtpole}_0}
\newc{\mbot}{m_b}
\newc{\vtb}{\widetilde V_{tb}}
\newc{\ths}{\theta_{\widetilde t}}
\newc{\mz}{m_Z}			\newc{\mw}{m_W}
\newc{\mhalf}{m_{1/2}}
\newc{\mgut}{M_X}
\newc{\ie}{{\it i.e.}}		\newc{\etal}{{\it et al.}}
\newc{\eg}{{\it e.g.}}		\newc{\etc}{{\it etc.}}
\newc{\hpm}{{H^\pm}}		\newc{\mhpm}{m_\hpm}
\newc{\stp}{{\widetilde t}}
\newc{\stl}{{\stp_L}}		\newc{\str}{{\stp_R}}
\newc{\stopl}{\stl}		\newc{\stopr}{\str}
\newc{\mstone}{m_\stone}	\newc{\stone}{{\widetilde t_1}}
\newc{\msttwo}{m_\sttwo} 	\newc{\sttwo}{{\widetilde t_2}}
\newc{\msbotl}{m_{\widetilde b_L}}
\newc{\msbotr}{m_{\widetilde b_R}}
\newc{\msbotone}{m_{\widetilde b_1}}
\newc{\msbottwo}{m_{\widetilde b_2}}
\newc{\mHonesq}{m^2_{H_1}}	\newc{\mHtwosq}{m^2_{H_2}}
\newc{\stau}{{\widetilde\tau}}
\newc{\gev}{{\rm\,GeV}}		\newc{\tev}{{\rm\,TeV}}
\newc{\mchone}{m_{\chone}}	\newc{\chone}{\chi_1^\pm}
\newc{\mchtwo}{m_{\chtwo}}	\newc{\chtwo}{\chi_1^\pm}
\newc{\mneone}{m_{\neone}}      \newc{\neone}{\chi_1^0}

\begin{titlepage}
\begin{flushright}
{\large
UM-TH-94-23\\
July 1994\\
}
\end{flushright}
\vskip 2cm
\begin{center}
{\Large\bf Implications of $\bf\Gb$
for Supersymmetry Searches and Model-Building}
\vskip 1cm
{\Large
James D.~Wells\footnote{E-mail: {\tt jwells@umich.edu}},
Chris Kolda\footnote{E-mail: {\tt kolda@umich.edu}},
and G.~L.~Kane\footnote{E-mail: {\tt gkane@umich.edu}}\\}
\vskip 2pt
{\large\it Randall Physics Laboratory, University of Michigan,\\ Ann Arbor,
MI 48109--1120, USA}\\
\end{center}
\vskip .5cm
\begin{abstract}
Assuming that the actual values of $\mtpole$ at FNAL
and of $\Gb$/$\Gh$ at LEP are within their current $1\sigma$ reported ranges,
we present a No-Lose Theorem for superpartner searches at LEP~II and an
upgraded Tevatron. We impose only two theoretical assumptions:
the Lagrangian is that of the Minimal Supersymmetric Standard Model (MSSM)
with arbitrary soft-breaking terms,
and all couplings remain perturbative up to scales $\sim10^{16}\gev$;
there are no assumptions about the soft supersymmetry
breaking parameters, proton decay,
cosmology, etc. In particular, if the LEP and FNAL values
hold up and supersymmetry is responsible
for the discrepancy with the Standard Model prediction of $\Gb$,
then we must have charginos and/or top squarks
observable at the upgraded machines (for LEP the superpartner threshold is
below $\sqrt{s}=140\gev$).
Furthermore, little deviation from the Standard Model is predicted within
``super-unified'' supersymmetry, so these models predict that the discrepancy
between experiment and the Standard Model prediction for $\Gb$ will fade with
time. Finally, it appears to be extremely difficult to find any unified MSSM
model, regardless of the form of soft supersymmetry breaking, that can explain
$\Gb$ for large $\tanb$; in particular, no model with $t-b-\tau$
Yukawa coupling unification appears to be consistent with the experiments.
\end{abstract}
\end{titlepage}
\setcounter{footnote}{0}
\setcounter{page}{2}
\setcounter{section}{0}
\setcounter{subsection}{0}
\setcounter{subsubsection}{0}

\section{Introduction}

Recent results from Fermilab~\cite{fnaltop}
indicate that the top quark is rather heavy ($\mtpole=174\pm 17$ GeV), while
recent results from LEP~\cite{leprb}
indicate that $\rb\equiv\Gb/\Gh$ might be inconsistent
(up to $2.5\sigma$) with such a heavy top.  In this letter we
assume that both measurements are correct, and point
out the rather powerful implications this has if nature is supersymmetric.

First, we give a brief discussion of the Standard Model (SM) prediction
for $\rb$ which is approximately $2$ to $2.5\sigma$ away from
the latest experimental
measurement.  We then consider the effect of supersymmetry (SUSY) on this
process.  In particular, we consider both the MSSM and the popular
``super-unified'' approach to supersymmetric model building.
We demonstrate a No-Lose Theorem for discovery of
superpartners at coming collider upgrades given $1\sigma$ experimental bounds
on $\rb$ and $\mtpole$, and a very minimal set of theoretical assumptions.

In considering the question of the $\rb$ discrepancy,
one can take either of two attitudes. Perhaps the $\rb$ measurement is finally
the one which directly demonstrates the existence of physics beyond the
Standard Model; if so, its implications for the discovery of supersymmetry
are dramatic (assuming SUSY is the origin of the deviation). Yet, as
we will discuss in this paper, the ``super-unified'' SUSY models
produce values for $\rb$ near those of the SM, not large enough to
explain the $\rb$ discrepancy.
Thus even if SUSY is the correct theory, it is not unlikely that the
measurements of $\rb$ will approach the SM expectation as systematic
effects are more fully understood.

\section{The Standard Model Prediction for $\bf\Gb$}
\bigskip

The ratio $R_b$ is very sensitive to vertex corrections involving
a heavy top quark. Within the Standard Model these
corrections are negative and grow like $m_t^2$.
(This can be seen best in the 't~Hooft-Feynman gauge where the
$\phi^+\bar t b$ coupling is proportional to $m_t$.)
On the other hand, the $m_t$-dependent
oblique corrections are to a good approximation universal and therefore
largely cancel out in the $R_b$ ratio.  This effectively isolates
the vertex corrections, thereby providing both an excellent test
of the SM's self-consistency and a place to search
for new physics beyond the reach of current experiments.

\begin{figure}
\centering
\epsfysize=3in
\hspace*{0in}
\epsffile{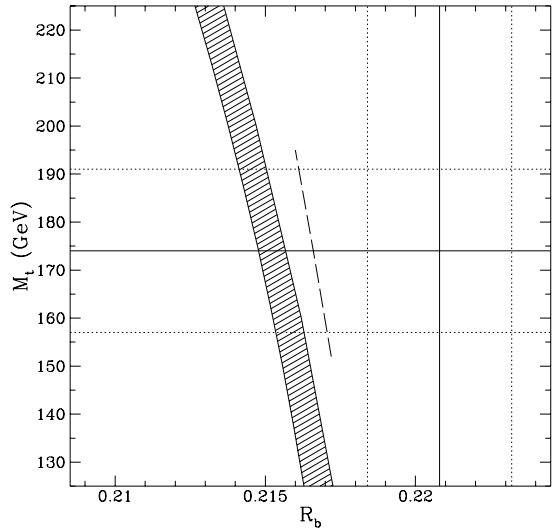}  
\caption{The SM prediction for $R_b$.  The
experimental central values
of $R_b$ and $\mtpole$ (along with their $1\sigma$ bounds) are designated by
straight solid (and dotted) lines. The shaded region
represents the SM prediction for $\rb$ from Z{\O}POLE, allowing for
the $b$-quark mass to vary within the range $4.5\leq\mbpole\leq5.3\gev$.
The curved dashed line represents the ``line of
closest approach'' for the CMSSM to the experimental $\rb$ value.}
\label{rbos}
\end{figure}

Using the program Z{\O}POLE~\cite{z0pole} we have calculated $R_b$ in the
on-shell scheme for $4.5\leq\mbpole\leq5.3\gev$ as a function
of the top quark pole mass, $\mtpole$.
In Fig.~\ref{rbos} we have plotted the experimental values (and
their $1\sigma$ ranges) for $\mtpole$ and
$\rb$. We also show the Z{\O}POLE calculation of the
SM prediction as the shaded region in the figure.
The SM prediction as quoted by LEP~\cite{leprb},
which is calculated by ZFITTER in the on-shell scheme, falls on the right
edge of the shaded region.
To be conservative we utilize the rightmost edge of the SM prediction
for $R_b$ in all our calculations and comparisons
since it lies closest to the the experimental measurement.

With an SM prediction in hand we now can compare with
LEP's measurement of $R_b$. We use $R_b=0.2208\pm 0.0024$~\cite{leprb}.
The uncertainty
includes the quoted statistical and systematic errors added
in quadrature.  One sees from Fig.~\ref{rbos} that
$R_b^{\rm theory}=0.2158$ for $\mtpole =174\gev$. This is approximately
2$\sigma$ away from the above quoted experimental measurement.
Disagreement of experiment and theory
for other values of $\mtpole$ can likewise be read off from
Fig.~\ref{rbos}. Since a heavy top forces $R_b$ downward, a rather large
positive contribution to $R_b$ from new physics clearly
is required to lift $R_b$ to the reported central value, given CDF's
measurement of a heavy top quark mass.

We understand of course that the measured $\rb$ may decrease as systematic
effects are better understood. However, if $\rb$ is the long-awaited
deviation from the SM that heralds the onset of new physics, its implications
for SUSY are dramatic. More precisely, the question we consider here is
the following:  What implications are there for supersymmetry
if the true value of $R_b$ {\em is} within one standard deviation
of the current measurement at LEP and the true value of $\mtpole$ {\em is}
within one standard deviation of the current measurement at FNAL?
Henceforth we premise the
true value of $R_b$ and $\mtpole$ to be within $R_b=0.2208\pm0.0024$ and
$\mtpole=174\pm17\gev$ respectively, and
investigate the consequences of this statement.  We find several
surprising results.

\section{Supersymmetry mass bounds from $\bf\Gb$}
\bigskip

In a supersymmetric
theory there are additional corrections to the $Zb\bar b$ vertex
which scale like $m_t^2$ from the charged Higgs--top quark loops
and the chargino--top squark loops. Furthermore, if $\tan\beta$ is
very large then, for example, the pseudoscalar Higgs ($A^0$)
coupling to $b\bar b$ is proportional to
$m_b\tan\beta$ and the bottom squark--$b$--neutralino
($\tilde b \bar b\chi^0$) coupling similarly is proportional to
$m_b/\cos\beta$.  For $\tan\beta \gsim 40$ these contributions
from ``neutral exchanges'' can be sizable. The supersymmetric
electroweak vertex corrections to the $Zb\bar b$ vertex at the
$Z$ pole have been analyzed by several
authors~\cite{denner,djouadi1,boulware,djouadi3}.
We primarily draw from the formulas of Ref.~\cite{boulware}
(see Appendix) in order to calculate the
supersymmetric contributions to $R_b$.

We will now investigate the
Minimal Supersymmetric Standard Model (MSSM), which we
define as the minimal (but most general)
$SU(3)\times SU(2)\times U(1)$ gauge invariant Lagrangian containing
the particle content of the SM, but which is supersymmetric
(up to soft-breaking terms) and $R$-parity conserving.

Despite the large number of unknown soft-breaking
parameters within the MSSM, any
individual process usually depends only on a small subset. Thus it is with
$\rb$, where only 12 unknown parameters of the MSSM enter. We will show that
the experimental values for $\rb$ may be used to constrain certain of these
parameters to ranges that can be easily probed in the near future.
Since the SM prediction for $R_b$ is below the experimental one,
the value which becomes most important to us for the present analysis
is $\rbmax$, the maximum value which
$\rb$ can take given any set of inputs and constraints.

The independent parameters which enter the calculation of $\rb$
are $\tanb$, $M_1$, $M_2$, and $\mu$ (from which one gets the chargino
and neutralino masses and mixings), the physical top and bottom squark
masses and their mixing angles ($m_{\tilde t_{1,2}}$, $m_{\tilde b_{1,2}}$,
$\theta_{\tilde b}$, and $\theta_{\tilde t}$), the pseudoscalar Higgs mass,
and $\vtb$ (the super-CKM angle between the bottom and top
squarks).  In theory, determining $\rbmax$ for any given set of assumptions
is very difficult, for it will be a complicated function of all the free
parameters which enter the process. Luckily, we can separate the dependences
on many of the parameters and work with them independently.
For example, $\rb(\vtb)$ is always maximal when $|\vtb|=1$, regardless of the
values of the other parameters.

Some parameters are {\em almost} separable. The top squark mixing parameter,
$\ths$, tends to maximize $\rb(\ths)$ for values which are small and negative.
Yet the choice $\ths=0$ always produces a near-maximal $\rb$, up to
corrections of order $0.01\sigma$, far too small for us to worry about here.
Having now chosen $\ths=0$ one then finds that $\sttwo(=\stopl)$
decouples and its mass is no longer one
of the parameters on which $\rb$ depends for low to intermediate $\tanb$.
Also in this region of $\tanb$, the bottom squark--neutralino
contributions decouple completely, leaving $\rb$ independent
of $m_{\tilde b_{1,2}}$, $\theta_{\tilde b}$, and $M_1$.

Finally, $\rb$ is most strongly dependent on the masses of the light
chargino(s), light top squark(s), and for large $\tanb$, neutral Higgs bosons.
Lighter top squarks simply give larger
contributions to $\rb$. The chargino contributions are
much more complicated since they induce a local maximum of $\rb$
right at $\mchone=\sqrt{s}/2=\mz/2$ where $\rb$
can be very large. Unfortunately, current bounds on the chargino mass fall
right at $\mz/2$; were these mass bounds a few GeV higher, much stricter
bounds, for example
on $\mstone$ and $\tanb$, could be determined. This will be
demonstrated explicitly later. Finally, the dependence on $\mA$ changes sign
as $\tanb$ increases, weakly favoring large $\mA$ (actually, large $\mhpm$)
for small to moderate $\tanb$, but strongly favoring small $\mA$ for
large $\tanb$. One has in total twelve parameters, though only seven
of them are relevant at low to intermediate values of $\tanb$; with any fewer
we lose complete generality.

\subsection{The No-Lose Theorem}
\bigskip

One may summarize the nature of our No-Lose Theorem thusly: SUSY is a
decoupling theory. That is, if SUSY is to make measurable contributions
to the radiative corrections of SM processes, the mass scale of the SUSY
partners must be near to the scale of the SM physics being studied. Only
for masses of spartners near to the current experimental limits can one
expect to notice their effects through loops in SM diagrams. As these masses
increase, the SM prediction for any given quantity is again realized. Thus,
if we are to explain the discrepancy between the experimental and theoretical
values for $\rb$ using SUSY, the SUSY mass scale cannot be large.

That said, it remains to be seen what kind of numerical bounds can be placed
on the masses of the sparticles entering into the SUSY contributions to
$\rb$. The final result of such an analysis yields this: {\em If the
discrepancy in $\rb$ is to be explained by the MSSM, then direct observation
of superpartners at LEP~II or an upgraded Tevatron will occur.} Strict bounds
may be placed on chargino and top squark masses if SUSY is to explain
the value for $\rb$: $\mchone<85\gev$ and $\mstone$ or $\msbotone<165\gev$.
Stronger bounds can also be found. For example, for $\tanb\lsim30$,
$\min(\mchone,\mstone)<65\gev$. Under additional
constraints, bounds exist also on $\mA$, $\mneone$,
and $\tanb$. This theorem holds
under the following set of assumptions which we will examine below:
{\em (i)} the true value for $\rb$ is within $1\sigma$ of quoted LEP
measurements, {\em (ii)} the true value for $\mtpole$ is within $1\sigma$ of
quoted CDF measurements, {\em (iii)} contributions from the MSSM are
responsible for the difference between the actual and theoretical SM values,
{\em (iv)} the Yukawa couplings of the MSSM remain perturbative up to
scales $\sim10^{16}\gev$ (a so-called perturbatively valid theory),
and {\em (v)} various experimental lower bounds on sparticle masses from
direct searches.

Let us examine these conditions. The very first condition is also the one
least trusted by us. Current LEP bounds on $\rb$
are $2$ to $2.5\sigma$ from theoretical expectations;
however, if the LEP measurement of $\rb$ migrates over time to the SM value,
our theorem will of course cease to be meaningful.

The second condition is important mostly for the $1\sigma$ lower bound
on $\mtpole$. In the SM as $\mtpole$ decreases, the prediction for $\rb$
increases such that the discrepancy between experiment and theory lessens.
Also, when coupled with the requirement of perturbative validity
(condition {\em(iv)} above), the lower bound on $\mtpole$ places a lower bound
on $\tanb$. This lower bound on $\tanb$ is
necessary for the existence of upper mass bounds (see below).

The third condition listed above is self-explanatory.

\begin{figure}
\centering
\epsfxsize=3.25in
\hspace*{0in}
\epsffile{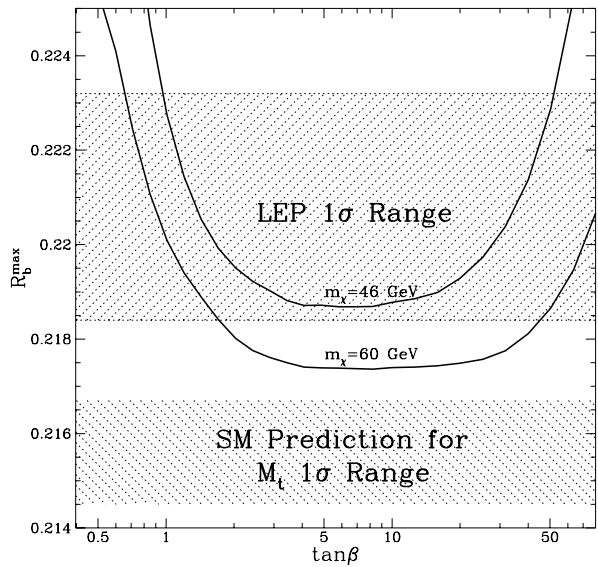}  
\caption{The dependence of $\rbmax$ on $\tanb$. The maximum possible value for
$\rb$ obtainable as a function of $\tanb$ is plotted for $\mchone=46\gev$
(upper line) and $\mchone=60\gev$ (lower line). The upper hatched region
is the experimental $1\sigma$ range for $\rb$, while the lower range
represents the SM range consistent with
the $1\sigma$ bounds for $\mtpole$.}
\label{alltanb}
\end{figure}

The fourth condition of perturbative validity is necessary in order to
place upper and lower bounds on $\tanb$. The requirement that the top,
bottom, and tau
Yukawa couplings remain perturbative up to a scale $\sim10^{16}\gev$
places limits on $\tanb$ of
\beq
\sin\beta>\frac{\mtpole}{\mtopmax}\quad{\rm and}\quad\tanb\lsim60
\eeq
where typically $190\lsim\mtopmax\lsim200\gev$. These limits are necessary
in order to gain mass bounds on the chargino and top squark. This can be
seen in Fig.~\ref{alltanb},
where the maximum value for $\rb$ is plotted against
$\tanb$ for two choices of $\mchone$. Clearly as $\tanb\to0$ and
$\tanb\to\infty$,
$\rb$ can become large even for heavy charginos and top squarks.

The fifth and final ``condition'' is the set of experimental lower bounds
that we apply to the masses of the MSSM. We take $\mchone>46\gev$ and
the rather conservative $\mstone>36\gev$~\cite{stops}. We take lower bounds on
the pseudoscalar mass from the non-observation of $Z\to h^0A^0$ (low
$\tanb$) and
$Z^0\to A^0b\bar b$ (high $\tanb$)
at LEP, with $A^0$ then decaying into $\tau^+\tau^-$. These bounds
are $\tanb$-dependent. For $1\lsim\tanb\lsim3$<, $\mA>20\gev$~\cite{aleph};
for $\tanb\gsim30$, $\mA>60\gev$~\cite{djouadi2}.

In the following sections we demonstrate how the bounds on chargino and
top squarks are determined through the set of five conditions above.

One should note that the bounds given above can be easily tightened.
In particular, if $\mtpole$ is found to be larger than its $1\sigma$
lower bound then our bounds will become stronger.
For $\mtpole=174$, the separate upper bounds on
$\mchone$ and $\mstone$ decrease to $63\gev$ and $77\gev$, with a
corresponding decrease in the minimum of the two.
Bounds on $\tanb$ will also exist if
values of $\mchone$ near to $46\gev$ are ruled out experimentally, as shown
by Fig.~\ref{alltanb}; there the lower line represents $\rbmax$ for
$\mchone>60\gev$, in which case we see that the entire region of
$2\lsim\tanb\lsim40$ is ruled out by the $1\sigma$ $\rb$ bounds.

\subsection{Low and Intermediate $\bf\tanb$ Region}
\bigskip

Consider first the case of low $\tanb$ (\ie, $\tanb\lsim5$). Here
one finds that $\rb$ increases monotonically as $\tanb$ decreases (see
Fig.~\ref{alltanb}). Therefore
$\rbmax(\tanb)$ is found at the boundary where $\sin\beta=\mtpole/\mtopmax$.
For intermediate $\tanb$ (\ie, $5\lsim\tanb\lsim30$), $\rbmax$ increases
with $\tanb$ but remains below the values obtained at very small
$\tanb$. Therefore, one may simplify the analysis in this region by only
considering the lowest possible value of $\tanb$ for a given $\mtpole$.

\begin{figure}
\centering
\epsfxsize=3.25in
\hspace*{0in}
\epsffile{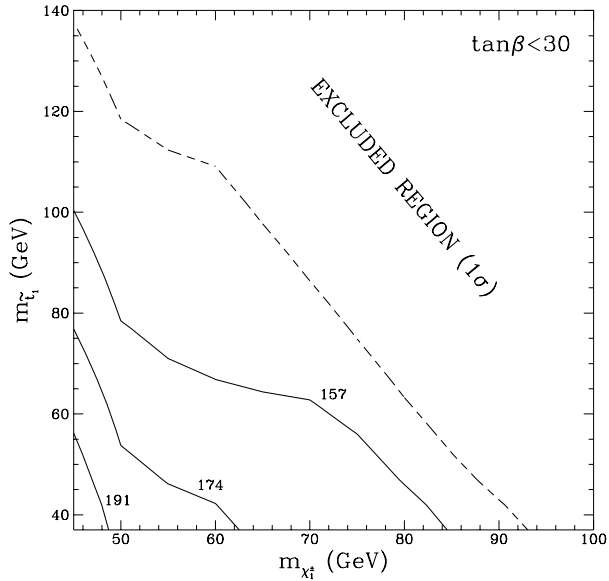}  
\caption{Upper bounds on ($\mchone,\mstone$) such that $\rb$ will fall
within $1\sigma$ of experiment, for $\tanb<30$.
Each solid line represents a different
value for $\mtpole$ (157, 174, and $191\gev$) such that only the region
below and to the left of the lines are consistent with the $\rb$
measurement, \ie, $\rb\geq0.2184$.
The region above and to the right of the $\mtpole=157\gev$ line is
excluded to $1\sigma$. In order to show how the limits are altered if
the experimental value for $\rb$ changes, we plot a dashed line
representing the
upper bounds for $\mtpole=174\gev$ and $\rb\geq0.2172$.
}
\label{ctlow}
\end{figure}

In Fig.~\ref{ctlow} we have plotted contours of $\rbmax=0.2184$
(the $1\sigma$ lower experimental bound) against
$\mchone$ and $\mstone$, for $\mtop=157$, $174$, and $191\gev$.
Recall that $\rbmax$ is the largest value that the theory can produce,
regardless of the values of the other parameters not shown in the plot.
To be conservative, we take $\mtopmax=205\gev$, thereby
allowing smaller $\tanb$ (and thus larger $\rbmax$) for a given top quark mass.
We emphasize that what is shown are the maximum values of
$\mstopone$ and $\mchone$ that can give $R_b$ within $1\sigma$
of its reported value. In the figure, the regions to the left and below
each line are compatible with the $1\sigma$ bounds on $\rb$; the regions
above and to the right are excluded.

One can read the central result of the No-Lose Theorem from this graph:
by combining the
$1\sigma$ bounds on the top quark mass from CDF and on $\rb$ from LEP, one
can place upper bounds of $85\gev$ on the mass of
the lighter chargino and $100\gev$ on the mass of the lighter top squark
in the low to intermediate $\tanb$ region. However, one or the other
must always be lighter than about $65\gev$. For
$\mtpole$ above $157\gev$, these limits become stronger. These bounds are to
our knowledge the strongest experimental bounds on the MSSM to date.
They do not depend on any scheme for
soft-breaking parameters, on constraints from
dark matter or proton decay, or any assumptions other than those few listed.
They imply
that with an upgrade in energy to $\sqrt{s}=130\gev$ at LEP a chargino or a
top squark must be
found, and with an upgrade in luminosity
at the Tevatron both the chargino and the top squark should be found.
Therefore, the ``$1\sigma$'' ranges of
chargino and top squark masses as shown in Fig.~\ref{ctlow}
could be completely probed if $\tanb\lsim 30$.

Further, if $\mchone\gsim60\gev$ and $\tanb\lsim40$, then $\tanb$ is restricted
to be close to 1. Therefore, the mass of the lightest Higgs
approaches zero at tree level, most of its mass then being due to radiative
corrections. Since we bound the lighter top squark mass,
the leading term in the
corrections to the Higgs mass (which goes as $m_t^4$) is determined up to
the value of the heavier top squark. Ignoring contributions due to
top squark mixing, one finds that for $\msttwo<1\tev$, $m_{h^0}<75\gev$;
likewise for $\msttwo<2\tev$, $m_{h^0}<80\gev$.
Thus in the low to intermediate $\tanb$ regime, $h^0$ may also be accessible
at LEP or FNAL.

\subsection{The High $\bf\tanb$ Region}
\bigskip

Large values of $\rb$ can also be attained for high $\tanb$.
Once again, it is at the perturbative edge ($\tanb\simeq60$)
that we obtain the largest $\rbmax$. Interestingly, it is also this
region of very high $\tanb$ that is motivated by $t-b-\tau$ Yukawa
unification within minimal SO(10) models.
In this region interactions proportional to $m_b\tanb$ become
comparable to and possibly more significant than the $m_t$ dependent
interactions.  Therefore, neutralino--bottom squark loops and neutral
Higgs--bottom quark loops must be considered.

Four of the
independent parameters of our model which could be ignored in the low
$\tanb$ region, namely the bottom squark masses and mixing angle, and $M_1$,
must now be included when we maximize $\rb$. This significantly complicates
the process and the demonstration of our theorem, so we again separate
those variables that can be taken independent from the others.
One simplification
would be to set $M_1\simeq\frac{1}{2}M_2$, as indicated by wide classes
of supergravity and superstring scenarios. We have taken this simplification
as well-motivated
and base most of our numerical results in the high $\tanb$ limit
on it; however, we have checked that perturbing the
ratio of $M_1$ to $M_2$ by an additional factor
of two in either direction only changes our calculations of $\rbmax$ by
less than $0.1\sigma$.

The bottom squark masses and mixing
provide another complication. Our calculations
indicate that $\rb$ is maximized when both bottom squarks are light, near to
their lower bound which we take to be $45\gev$. The squarks are then nearly
degenerate in mass and the mixing angle becomes arbitrary.

The Higgs sector behaves in the high $\tanb$ limit very differently than
in the opposite limit.
Unlike in the low $\tanb$ case, a light pseudoscalar Higgs yields an overall
large {\em positive} contribution to $\rb$. The contribution of the light
charged Higgs is still negative, but is overwhelmed by a large contribution
from neutral Higgs--bottom quark loops which increases with $\tanb$.
Therefore,
we must reintroduce the pseudoscalar mass as a variable when working
in this region. However the experimental constraint
that $\mA>60\gev$ for $\tanb>30$~\cite{djouadi2} leads to chargino bounds at
large $\tanb$ which are tighter than those at small $\tanb$.

How do discovery limits compare in the high $\tanb$ region? Most
significantly, one finds that $\mchone\lsim70\gev$, a tighter bound than
existed in the low $\tanb$ limit. However, it is not solely the
chargino--top squark loops that are responsible for this bound, but also
the neutralino--bottom squark loops. As the chargino mass becomes larger and
its contributions decouple, $\rbmax$ plummets towards the SM value,
thereby placing a bound on $\mchone$. Yet as $\tanb$ increases, the neutralino
contributions can be sizable, bringing $\rb$ back into agreement with
experiment. Thus there is really a bound on a combination of
the neutralino and chargino masses. But the appearance of the same $\mu$ in
both sectors means that bounds can be placed on each individually,
leading to the $70\gev$ for the
chargino given above. The comparable neutralino bound is then approximately
$67\gev$. If the $1\sigma$ lower bound of $\rb$ shifted to $0.2172$, then
the bounds on the chargino and neutralino would relax to $98$ and $95\gev$
respectively.

The same interplay we found between chargino and neutralino masses is of
course also found in the top and bottom squark sectors. Under the
assumption of nearly degenerate bottom squarks, one finds that either
a bottom {\em or} a top squark must be lighter than $165\gev$ ($85\gev$)
for $\tanb\leq60$ and $\mchone\geq45\gev$ ($\mchone\geq60\gev$),
a weaker bound than was found in the low $\tanb$ limit. But unlike the
case for the charginos and neutralinos, no individual bounds exist on
the top and bottom squarks alone.

Finally, there are also the contributions due to $A^0$--bottom quark loops.
Once again an upper bound can be placed on $\mA$ consistent with $\rb$
if $\mchone\geq60\gev$. Here we find that $\mA<95\gev$ is necessary. Thus,
light pseudoscalars are implied by $\rb$ for high $\tanb$, once
one constrains charginos to be slightly heavier than the current bound
of $46\gev$.

\section{Implications for the Super-Unified MSSM}
\bigskip

Much work has been completed recently by a number of groups on the
phenomenology of ``super-unified'' minimal SUSY (see Ref.~\cite{kkrw} and
references therein). These models are
constructed under the assumption of not only gauge coupling unification at
some high scale, but also unification of various soft mass parameters
in the MSSM Lagrangian (a common gaugino and a common scalar mass).
One then connects these high scale assumptions
to low-energy phenomenology through the renormalization group equations (RGE's)
of the parameters, under the condition that electroweak symmetry-breaking
occurs radiatively at scales $\sim\mz$.

In two previous works~\cite{kkrw,kkrw2},
the super-unified MSSM was assumed and a number of constraints stemming from
direct experimental searches for SUSY, CLEO bounds on ${\rm BR}(b\to s\gamma)$,
relic abundances of the lightest SUSY particle, etc. were assumed.
What remained of the
original parameter space of the super-unified models was called the
Constrained Minimal Supersymmetric Standard Model (CMSSM). The natural question
is, then, what ranges of values for $\rb$ are predicted for solutions
consistent with the CMSSM?  In fact, one finds that the constraints of the
CMSSM force $\rb$ to lie below approximately $0.2166$, some $1.5\sigma$ below
the reported $\rb$ for $\mtpole=174\gev$.
Fig.~\ref{rbos} shows the CMSSM's ``line of closest approach'' to the
experimentally measured value of $R_b$.  This closest approach
line does not enter the 1$\sigma$ area, and therefore CMSSM
cannot bring LEP's measurement of $R_b$ into agreement with FNAL's
measurement of $\mtpole$; conversely, one could say that the CMSSM predicts
$\rb$ below the line given in Fig.~\ref{rbos}.

Why is the CMSSM incapable of producing larger values for $\rb$?
Naively, one would expect contributions from supersymmetric masses
$\sim\mw$ to have a large contribution to the $Z\to b\bar b$
partial width just as the $W$ boson itself has.  As described earlier,
there exist interactions of the charginos with top squarks and quarks which
are large, proportional to the top Yukawa coupling ($h_t$),
and enhance $\rb$. Similar contributions exist at large $\tanb$ for the
neutral Higgs bosons with bottom quarks and squarks,
proportional to the $\tanb$-enhanced bottom Yukawa coupling. These interactions
could lead to $\rb$ consistent with experiment.
We now summarize the reasons that they do not.

The coupling of the top quarks to charginos and top squarks, proportional
to the large top Yukawa coupling, is given by
${\cal L}=h_t \bar t \tilde H^\pm \tilde t_{R}.$
In order to maximize the impact of this coupling,
we need a light top squark with a significant right-handed
component and a light chargino with a significant higgsino
component.  These are both difficult requirements for the CMSSM,
and they are quite impossible to satisfy simultaneously.

Within CMSSM, the choice of common scalar masses means that
$m_{\tilde t_R}$ is invariably smaller
than $m_{\tilde t_L}$ due to the running of the RGE's from
the GUT scale down to the weak scale.  Yet the resulting $\mstone$
is rarely smaller than $\mw$ unless we impose large mixing between the
$\stopl$ and $\stopr$ eigenstates. This in turn means that $\tilde t_1$
will have a significant $\stopl$ component which does not
couple to the charginos with $h_t$.

The lightest chargino is even more troublesome.
The chargino mass matrix depends on $M_2$, $\mu$ and $\tanb$.
With radiative breaking $\mu$ scales roughly
as max$\{m_{1/2},m_0\}$ and thus generally dominates the
chargino matrix (for a discussion, see Ref.~\cite{kkrw}).
This means that the charginos with significant higgsino components are
heavy, generally well above $\mw$,
and even though they couple with a factor of $h_t$,
these contributions will be kinematically suppressed.
Solutions with a chargino as light as $46\gev$ can be obtained,
but since they will be almost all $\widetilde W^\pm$
they do not couple as $h_t$.

When we try to simultaneously satisfy the
two requirements of a light $\stone\simeq\stopr$ and a light
$\chi^\pm\simeq\widetilde H^\pm$,
we find that the CMSSM is incapable of providing
a solution.  The two requirements actually push solutions in
different, incompatible directions.  Large mixing in the
top squark sector generally requires $\mu$ (or $m_0$) to
be large, while light higgsino-like charginos require
$\mu$ (thus $m_0$ also) to be small.

In the high $\tanb$ region, the additional contributions coming from the
pseudoscalar--bottom squark loops are likewise suppressed. The CMSSM generally
produces heavy bottom squarks and also heavy pseudoscalars, so that their
contributions will be small and cannot bring the CMSSM into agreement with
experiment.

For these reasons, and after much numerical work, we can conclude that
the CMSSM is incompatible (to $1.5\sigma$) with the experimental
measurements of $R_b$ and $\mtpole$.  Those who wonder how easy it
is to vary SUSY parameters in order to fit data should note this result.
Clearly, the CMSSM's ability to fit experiment
{\it for this particular observable}
is inextricably tied to the Standard Model's since the
predictions are essentially the same.

What of models which unify the gauge couplings but not the soft-breaking
parameters? One may use a ``bottom-up'' approach in building models at
the GUT scale, starting from the constraints of the low-energy theory.
Of particular interest is the case motivated by $t-b-\tau$ Yukawa coupling
unification in SO(10). Here one finds that large $\tanb$ is necessary,
which could be consistent with $\rb$ given the increase in $\rbmax$ that occurs
at very large $\tanb$ (see Fig.~\ref{alltanb}). However, the RGE's for the
soft masses can be analytically solved in this (pseudo-fixed point)
limit~\cite{carena}, and are consistent with a tree-level $\mA^2>0$ only if
$\mu^2\gsim 3M_{1/2}^2$. In this case, the light charginos/neutralinos
are predominantly wino/bino, not higgsino, and $\rb$ will approach the SM
value. Therefore, these SO(10)-type models appear to not be able to yield
$\rb$ within $1\sigma$ of experiment, regardless of the GUT-scale structure
of the soft-breaking terms.

\section{Conclusions}

We have explicitly demonstrated a type of No-Lose Theorem for direct
observation of SUSY partners at either LEP~II or upgraded Tevatron. The
central assumption for this theorem is that the actual values of $\rb$
and $\mtpole$ lie at or above their $1\sigma$ lower bounds.

By coupling these measurements to the requirement of perturbative validity of
the Yukawa couplings up to a GUT-like scale, we derived bounds on charginos and
top squarks leading one to conclude that one or both
should be directly observed
at LEP~II with $\sqrt{s}>140\gev$, and at an upgraded (in luminosity)
Tevatron, for any SUSY breaking and any (perturbative) $\tanb$.
We also pointed out the existence of
interesting bounds on $\tanb$ if the lower bound on charginos could be
pushed up a little from its current value. Of course, these bounds
may never be saturated. Other observables, such as the $\rho$-parameter,
the forward-backward $b$-asymmetry ($A^b_{\rm FB}$), or
$B^0-\bar{B^0}$ mixing ($\chi_B$) could significantly tighten our upper bounds;
we are currently examining this issue. However, we have checked that full
spectra can be found which are consistent with both the experimental value for
$\rb$ and a cosmological relic density of $\Omega\simeq1$.

Finally, we explained the inability of the simplest class of
super-unified models to explain the $\rb$ discrepancy. We found that
these models are strongly decoupled from the decay of $Z\to b\bar b$
despite their often low mass scales, so
that their prediction for $\rb$ bears little difference from that of the
SM. We are currently exploring the question of
what kinds of GUT models can be found which {\em are} consistent with $\rb$.
In particular, one wishes to know what hierarchies of soft-breaking
masses are required at the GUT scale to replace the assumption of common
masses. Surprisingly, we have found that it is
extremely difficult to construct a unified theory for any set of soft-breaking
parameters if $\tanb$ is in the range required by unification of all
third generation Yukawa couplings, as in SO(10) models.
The solution to this problem may lead to progress in
understanding the mechanism of soft SUSY breaking and to the breaking
of gauge symmetries at the GUT scale, assuming the experimental
$\rb$ discrepancy persists over time.

\section*{Acknowledgements}
This work was supported in part by the U.S. Department of Energy and
and the Texas National Research Laboratory Commission.
Much of the computational work was done with the aid of
LERG-I~\cite{lergi} and Z{\O}POLE~\cite{z0pole}. We would also like to
thank M.~Beneke, A.~Blondel, M.~Einhorn,
K.~Riles, O.~Rind, L.~Rolandi, G.~Ross, L.~Roszkowski,
R.~Stuart, and D.~Treille for useful discussions and commnuications.

\section*{Appendix}

Supersymmetric corrections to the $Zb\bar b$ vertex
are $m_t$ and $m_b\tan\beta$ dependent, and can be quite large.
In this appendix we present the calculation of the shift in $R_b$
due to supersymmetric contributions from charged Higgs--top quark loops,
chargino--top squark loops, and neutralino--bottom squark loops.
Our calculations are in agreement with those of Ref.~\cite{boulware},
and in this appendix we follow the notation of this reference as closely
as possible.
However, we attempt to clear up possible confusion by presenting the equations
strictly in terms of
ordinary Passarino-Veltman functions with full arguments.

\begin{figure}
\centering
\epsfxsize=2.75in
\hspace*{0in}
\epsffile{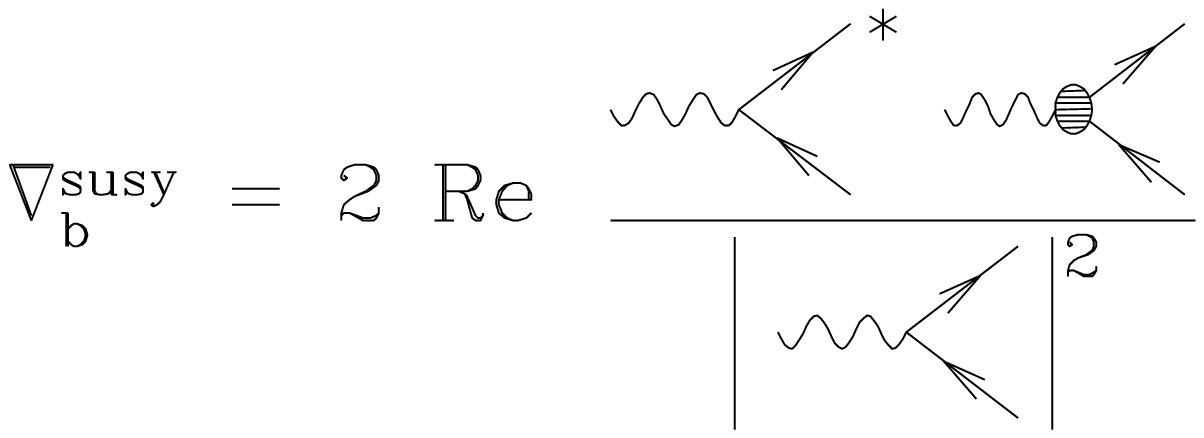}  
\caption{Feynman diagram representation of $\nabla_b^{\rm susy}$.}
\label{nabla:diagram}
\end{figure}

The expansion of $R_b$ can be separated into SM contributions
and supersymmetric contributions~\cite{boulware}:
\bea
R_b &= & \Gb/\Gh \nonumber \\
    &= & R_b^{\rm sm}(m_t,m_b)+R_b^{\rm sm}(0,0)[1-R_b^{\rm sm}(0,0)]
  [\nabla^{\rm susy}_b(m_t,m_b)-\nabla^{\rm susy}_b(0,0)].
\eea
$R_b^{\rm sm}(0,0)$ is the SM prediction
of $R_b$ for massless top and bottom quarks which we take to
be 0.220.  The value for $\nabla^{\rm susy}_b(m_t,m_b)$ is the
sum of one-loop interferences with the tree graph divided
by the squared amplitude of the tree graph as shown
in Fig.~\ref{nabla:diagram}.

A convenient parameterization of the supersymmetric vertex corrections
$\nabla^{\rm susy}_b(m_t,m_b)$ is
\beq
\nabla_b^{\rm susy}(m_t,m_b)=
    \frac{\alpha}{2\pi\sin^2\theta_W}\frac{v_L F_L+v_R F_R}{(v_L)^2+(v_R)^2},
\eeq
where
\beq
v_L=-\frac{1}{2}+\frac{1}{3}\sin^2\theta_W,~~v_R=\frac{1}{3}\sin^2\theta_W.
\eeq

Loops involving the charged Higgs with the top quark yield the following
contributions to $F_L$ and $F_R$:
\bea
F_{L,R}^a &= & S_1(m_t,\mhp) v_{L,R} \lambda^2_{L,R} \\
F_{L,R}^b &= &[S_2(m_t,\mhp,m_t) v^{(t)}_{R,L}
              +m_t^2 S_3(m_t,\mhp,m_t) v^{(t)}_{L,R}] \lambda^2_{L,R} \\
F_{L,R}^c &= &S_4(\mhp,m_t,\mhp) (\frac{1}{2}-\sin^2\theta_W) \lambda^2_{L,R}
\eea
where
\bea
v^{(t)}_L &=&
\frac{1}{2}-\frac{2}{3}\sin^2\theta_W,~~v^{(t)}_R=-\frac{2}{3}\sin^2\theta_W \\
\lambda_L &=&
\frac{m_t\cot\beta}{\sqrt{2}m_W},~~\lambda_R=\frac{m_b\tan\beta}{\sqrt{2}m_W}.
\eea
Loops involving charginos and top squarks yield the following contributions
to $F_L$ and $F_R$:
\bea
F_{L,R}^a&=&\sum_{i,j} S_1(\xci,\stj) v_{L,R}
             \Lambda_{ji}^{*L,R} \Lambda_{ji}^{L,R} \\
F_{L,R}^b&=&\sum_{i,j,k}S_4(\sti,\xck,\stj)
             (\frac{2}{3}\sin^2\theta_W\delta_{ij}-\frac{1}{2}T^*_{i1} T_{j1})
             \Lambda_{ik}^{L,R} \Lambda_{jk}^{*L,R} \\
F_{L,R}^c&=&\sum_{i,j,k}[S_2(\xci,\stk,\xcj) O_{ij}^{R,L}+\xci\xcj
S_3(\xci,\stk,\xcj) O_{ij}^{L,R}] \Lambda_{ki}^{L,R} \Lambda_{kj}^{*L,R}
\eea
where
\bea
\Lambda_{ij}^L &= &T_{i1}V^*_{j1}
                  -\frac{m_t}{\sqrt{2}m_W\sin\beta}T_{i2}V^*_{j2} \\
\Lambda_{ij}^R &= &-\frac{m_b}{\sqrt{2}m_W\cos\beta} T_{i1}U_{j2} \\
O_{ij}^L &=& -\cos^2\theta_W\delta_{ij}+\frac{1}{2}U^*_{i2}U_{j2} \\
O_{ij}^R &=& -\cos^2\theta_W\delta_{ij}+\frac{1}{2}V_{i2}V^*_{j2}.
\eea
We follow the conventions of reference~\cite{haber} for the chargino
mixing matrices $U$ and $V$ defined in the
$\{\widetilde W^+,\widetilde H^+\}$ basis.

Loops involving neutralinos and bottom squarks
yield the following contributions to $F_L$ and $F_R$:
\bea
F_{L,R}^a&= &\sum_{i,j}S_1(\xni,\sbj) v_{L,R}
              \Lambda_{ji}^{*L,R} \Lambda_{ji}^{L,R} \\
F_{L,R}^b&= &\sum_{i,j,k}S_4(\sbi,\xnk,\sbj)
            ( \frac{1}{2}B^*_{i1}B_{j1}-\frac{1}{3}\sin^2\theta_W\delta_{ij})
            \Lambda_{ik}^{L,R} \Lambda_{jk}^{*L,R}  \\
F_{L,R}^c&= &\sum_{i,j,k}[S_2(\xni,\sbk,\xnj) O_{ij}^{R,L}
               +\xni\xnj S_3(\xni,\sbk,\xnj) O_{ij}^{L,R}]
	       \Lambda_{ki}^{L,R} \Lambda_{kj}^{*L,R}
\eea
where
\bea
\Lambda_{ij}^L &= &\frac{1}{\sqrt{2}}
                (\frac{1}{3}N^*_{j1}\tan\theta_W-N^*_{j2}) B_{i1}
                -\frac{m_b}{\sqrt{2}m_W\cos\beta}N^*_{j3}B_{i2} \\
\Lambda_{ij}^R &= & \frac{\sqrt{2}}{3}\tan\theta_W N_{j1} B_{i2}
                   -\frac{m_b}{\sqrt{2} m_W \cos\beta} N_{j3} B_{i1} \\
O_{ij}^L &=& \frac{1}{2}N^*_{i3} N_{j3}-\frac{1}{2}N^*_{i4}N_{j4} \\
O_{ij}^R &=& -O_{ij}^{*L}.
\eea
Again, we follow the conventions of reference~\cite{haber},
and define the neutralino mixing matrix $N$ in the
$\{\widetilde B,\widetilde W^3,\widetilde H_d,\widetilde H_u\}$ basis.
Note that this convention is different from that of
Ref.~\cite{boulware} which chose to diagonalize the neutralinos in the
$\{\widetilde B,\widetilde W^3,\widetilde H_u,\widetilde H_d\}$ basis,
thereby interchanging $N_{i3}$ and $N_{i4}$ in the above equations.

The $S_n$ functions are defined in terms of the non-divergent parts
of Passarino-Veltman scalar integral functions~\cite{passarino}:
\bea
S_1(m_1,m_2) &= & B_1(-m_b^2;m_1^2,m_2^2) \\
S_2(m_1,m_2,m_3) &= & {\textstyle-\frac{1}{2}}
		      +[2C_{24}-m_Z^2C_{12}-m_Z^2C_{23}]
                      (-m_b^2,-m_b^2,-m_Z^2;m_1^2,m_2^2,m_3^2) \\
S_3(m_1,m_2,m_3) &= & C_0(-m_b^2,-m_b^2,-m_Z^2;m_1^2,m_2^2,m_3^2) \\
S_4(m_1,m_2,m_3) &= & -2C_{24}(-m_b^2,-m_b^2,-m_Z^2;m_1^2,m_2^2,m_3^2).
\eea
All logs encountered in the scalar integrals are made dimensionless by
inserting the 't~Hooft mass $\mu_R$. Note that our $S_2$ function contains
$C_{12}$. In Ref.~\cite{boulware}, the calculation of $\rb$ is presented
in terms of reduced Passarino-Veltman functions~\cite{ahn}. However,
the translation there of the ${\bf c}_6$ reduced Passarino-Veltman function
leaves one to believe that $S_2$ contains
$C_{11}$ rather than the correct $C_{12}$.

\vfill\eject

\end{document}